# Photonic bandgap properties of hyperuniform systems self-assembled in a microfluidic channel

**Lily Traktman,**[1] **Bowen Yu,**[1] **Isa Vasquez,**[1] **Stanislav Ospov,**[1] **Remi Dreyfus,**[2,3] **and Weining Man**[1,*]

[1] *Department of Physics and Astronomy, San Francisco State University, San Francisco, CA 94132, USA*
[2] *Institut Interdisciplinaire d'Innovation Technologique (3IT), Université de Sherbrooke, 3000 Boulevard de l'université, Sherbrooke, J1K OA5 Québec, Canada*
[3] *Laboratoire Nanotechnologies Nanosystèmes (LN2) - IRL3463, CNRS, Université de Sherbrooke, Université Grenoble Alpes, École Centrale de Lyon, INSA Lyon, Sherbrooke, J1K 0A5 Québec, Canada*
*\*weining @sfsu.edu*

**Abstract: Traditional self-assembly methods often rely on densely packed colloidal crystalline structures and have inherent limitations in generating materials with isotropic photonic bandgaps (PBG). This study explores the photonic properties of materials structured according to hyperuniform disordered patterns (HUDS) generated via a hydrodynamic process in a microchannel. This research employs simulations to characterize optical bandgaps and determine the minimum dielectric contrast required for PBG formation in structures based on the templates experimentally formed under various conditions during the hydrodynamic process. The optimal conditions in the hydrodynamic process for realizing PBG have been identified. The findings offer a promising avenue for the large-scale production of isotropic photonic bandgap materials.**

## 1. Introduction

Photonic bandgap materials (PBM) are of significant interest in optical applications as they allow control of light propagation without loss [1]. To date, most PBMs have been fabricated using top-down approaches [2], such as electron beam lithography or photolithography, which are costly and impractical for large-scale production. For the past 30 years, large-scale production of such materials using a bottom-up approach has remained challenging. The most common approach currently relies on colloidal crystalline templates [3–5] around which a material of high refractive index is solidified [6]. However, this approach has two major limitations. First, the crystalline structure obtained from colloids is always densely packed, which is not suitable for the opening of a photonic bandgap. Second, the crystalline structures are anisotropic. They have specific orientational symmetry choices that limit the functional defect designs [7]. Furthermore, these structures can exhibit poor crystallinity; defects in colloidal crystals ultimately influence light transport within the material [8]. One way to circumvent these two major drawbacks is to develop other non-crystalline, self-assembled isotropic structures that possess photonic bandgaps.

Non-crystalline structures have gained renewed interest. Simulations have shown that, under well-controlled conditions, foams, for instance, can form a potential class of materials templates for the formation of 2D photonic bandgaps for the TE polarization [6] or even a complete photonic bandgap in 3D [9,10]. In general, a novel class of non-crystalline structures that can potentially be used as PBM was discovered: the hyperuniform disordered structures (HUDS) [11]. Hyperuniformity is assessed in Fourier space, using the structure factor of the material $S(q)$, where $q$ is the wavevector. The system is hyperuniform if $S(q)$ vanishes as $q$ approaches zero [11,12]. Hyperuniformity is shown to be related to the photonic bandgap properties of disordered 2D systems. Indeed, numerical simulations [13] and experiments [14,15] have shown that materials made of dielectric cylinders or walls located at the vertices or edges of the Voronoi cell derived from a stealthy hyperuniform pattern [16] possess sizeable photonic bandgaps for TM or TE polarizations. Combining the cylinders and network produces a complete bandgap for both polarizations [13,14]. Using 3D printing technology, Muller *et al.* [2] demonstrated that 3D HUDS with the right index mismatch exhibit a 3D photonic bandgap.

So far, there have only been a few examples of processes that drive matter to self-assemble into HUDS [9,17–25]. This has been a significant challenge until Weijs *et al.* [17] discovered that a droplet emulsion driven back and forth in a microchannel can self-organize into a HUDS configuration. Within a microchannel, droplets are periodically displaced with an average displacement amplitude of $\Delta$. It was observed that below a critical value of the displacement amplitude $\Delta^*$, the assembly organizes into a HUDS, whereas above $\Delta^*$, the emulsion droplets assemble into a non-hyperuniform disordered state. The transition at $\Delta^*$ is discontinuous, which is a feature of a first-order dynamic transition. While the discontinuous transition at $\Delta$ has been characterized, it remains uncertain whether the resulting HUDS can serve as effective templates for PBM formation [26].

In this article, we have investigated the optical properties of the different HUDS configurations obtained from the hydrodynamic experiments briefly described above. For this purpose, we consider the structures that can be obtained by solidifying materials with different refractive indices (or relative permittivities) using the hydrodynamics-experiment-generated emulsion configurations as templates. For each configuration obtained at different mean amplitudes of displacement $\Delta$ in the microfluidic channel, we compute the fully vectorial eigenmodes of Maxwell's equations with periodic boundary conditions using the free software MPB [27]. Additionally, we performed finite-difference time-domain (FDTD) simulations [28] using the open-source software package MEEP [29] to characterize material optical bandgaps. We then determine the minimum value $\varepsilon_{r,min}$ of the dielectric contrast $\varepsilon_r$ required for opening a photonic bandgap. $\varepsilon_r$ is defined as $\varepsilon_r = \varepsilon_{high}/\varepsilon_{low}$, where $\varepsilon_{high}$ is the relative permittivity of the material located at the droplets and $\varepsilon_{low}$ is considered to be 1.000 for air after the liquid phase around the solidified structure is removed. We demonstrate that right at the transition $\Delta^*$, the droplets self-organize into an optimal configuration, minimizing the dielectric contrast $\varepsilon_{r,min}$ required for the opening of a photonic bandgap. We observe a similar first-order phase change for $\varepsilon_{r,min}$ below and above the critical amplitude $\Delta^*$. The predicted values of the required dielectric contrast at the transition can be achieved using regular existing materials, suggesting a possible new approach to producing large-scale templates for making PBM. Moreover, the corresponding gap width at a given dielectric contrast $\varepsilon_r$ can be tuned by choosing different displacement amplitudes $\Delta$ in the hydrodynamic process.

2. **Methods of generating and characterizing the hyperuniform and non-hyperuniform disordered structures**

The templates of the disordered structures used in this study are based on the experimental results presented by Jeanneret *et al.* [26]. In that article, the authors present an experiment where an assembly of oil droplets is first generated in a microfluidics channel. Once this assembly is settled in the microchannel, a periodic flow is established, displacing the droplet longitudinally back and forth. Though the flow is longitudinal, hydrodynamic interactions between the droplets induce lateral displacements that are readily observable under optical microscopy. As explained in detail in [26], in the regime of low-Reynolds numbers, where viscosity dominates over inertia, periodic flows are expected to be reversible, meaning that a droplet, when displaced back and forth, is expected to come back to its same initial position. Jeanneret *et al.* [26] show that below a certain amplitude of displacement $\Delta^*$, the flow is reversible, whereas above $\Delta^*$, the flow becomes irreversible. A so-called fidelity function is defined as a measure of how much the flow is reversible. The authors show that the fidelity function experiences a discontinuity at $\Delta^*$: the transition between the reversible (below $\Delta^*$) and the irreversible regime (above $\Delta^*$) is actually a first-order dynamic transition. In a later article by Weijs *et al.* [17], based on the data obtained by Jeanneret *et al.* [26], it was discovered that not only does the fidelity function exhibit a discontinuity, but also that the structure factor experiences a sudden change at $\Delta^*$ as shown in Fig. 1. Below $\Delta^*$ (blue curves in Fig. 1) the structure factor dramatically decreases as *q* becomes smaller than the *q* values of the main peak. This is the characteristic structural behavior of a hyperuniform disordered structure, suggesting that, in a window of a particular size, the number fluctuation of particles normalized by the number of particles tends to decrease as the window size increases, while at $\Delta^*$(green curve in Fig. 1) and above $\Delta^*$(yellow curves in Fig. 1), the structure factor reaches a plateau as q vanishes. This is typical of point distributions with large density fluctuations, even at large distances, analogous to fluctuations that random distributions exhibit [11,21,30].

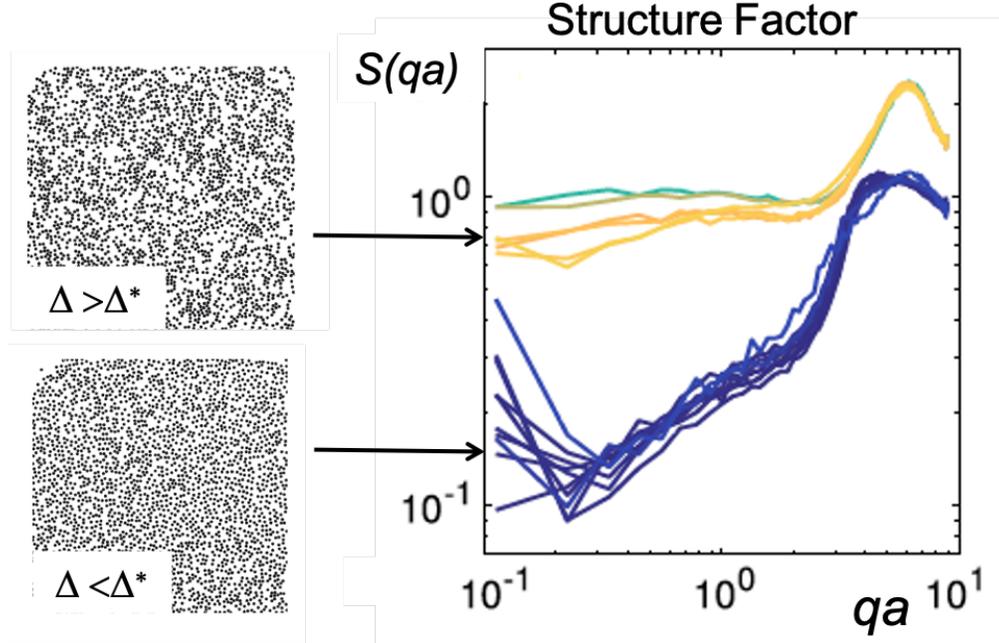

*Figure 1: Example of two droplet configurations obtained above the transition (top left) and right below the transition (bottom left), along with the corresponding structure factors [17]. Blue curves ($\Delta < \Delta^*$), yellow curves ($\Delta > \Delta^*$), green curves (right above $\Delta^*$)*

This previous study provides us with a unique example of an experimental approach, which allows us to tune and control the structural organization of a large population of droplets. Droplet populations organize in isotropic disordered states without any translation or rotational invariance. However, among these disordered states, some are hyperuniform and less "disordered" than others, since their structure factor vanishes at small *q* values. They also exhibit better local similarities in terms of densities and neighbor distributions, while local similarities have been found to be crucial for PBG formation [31]. Based on this observation, one may wonder if this transition correlates with a transition in the optical properties of structures that can be made using these experimentally obtained distributions of droplets.

To investigate the photonic properties of such structures, we extract the position of each droplet from the experimental data and consider that the position of each droplet can be used as a base to grow a cylinder of different materials with a radius *r* similar to the droplet radius. The cylinders will be tall and perpendicular to the plane of the 2D droplet pattern. We then assign a dielectric permittivity $\varepsilon_{high}\varepsilon_0$ to each cylinder. Finally, we assume that the liquid surrounding the droplets has been replaced by air and assign the permittivity in air ($\varepsilon_{air}\varepsilon_0$) to the continuous phase. The dielectric contrast $\varepsilon_r = \varepsilon_{high}\varepsilon_0/\varepsilon_{air}\varepsilon_0 = \varepsilon_{high}/\varepsilon_{air} = \varepsilon_{high}$ is the ratio of the dielectric constants between the solid and air parts (the black vs. white parts in Fig. 1). The dielectric contrast determines the refractive-index mismatch and significantly impacts the optical properties of the structure.

Through simulations, we analyze wave propagation in this medium, with a particular focus on the potential formation of a TM photonic bandgap. Simulations were performed with the finite-difference time-domain (FDTD) method [28], using an open-source software package, MEEP [29], to find out the transmission spectrum through such a medium. More importantly, we aim to identify the conditions for opening a forbidden frequency region with the absence of either propagating or localized states [15]. Fully transverse vectorial eigenmodes of Maxwell's equations with periodic boundary conditions on a large supercell were computed by preconditioned conjugate-gradient minimization of the block Rayleigh quotient in a plane wave basis, using a freely available software package called MPB [27], assuming the structure is infinitely long in the direction perpendicular to the 2D plane. A supercell containing over 600 cylinders is employed for each of the MPB simulations. The convergence of the results for larger supercell sizes has been confirmed.

### 3. Results and discussions

3.1 Simulations of band structures and field distribution

We first focus on a configuration experimentally obtained right below the critical displacement $\Delta^*$. A square-shaped super-cell consisting of N=631 particles is cropped from this oil-droplet configuration. As mentioned in Section 2, we assign different dielectric constants for the isolated phase (cylinders $\varepsilon_{high}$) and the continuous phase (voids $\varepsilon_{air}$=1). Two extreme cases are first considered, one of low dielectric contrast (case 1, $\varepsilon_r$=2) and one of large dielectric contrast (case 2, $\varepsilon_r$=12). The results are shown in Figure 2 and Figure 3, respectively. For the simulation, the average spacing between the particles $a$ is defined as the square root of the quotient of the super-cell size $A$ divided by the number of cylinders N. The unit length is defined as: $a = \sqrt{\frac{A}{N}}$ and the frequency result is in the unit of (c/a), where c is the speed of light in vacuum.

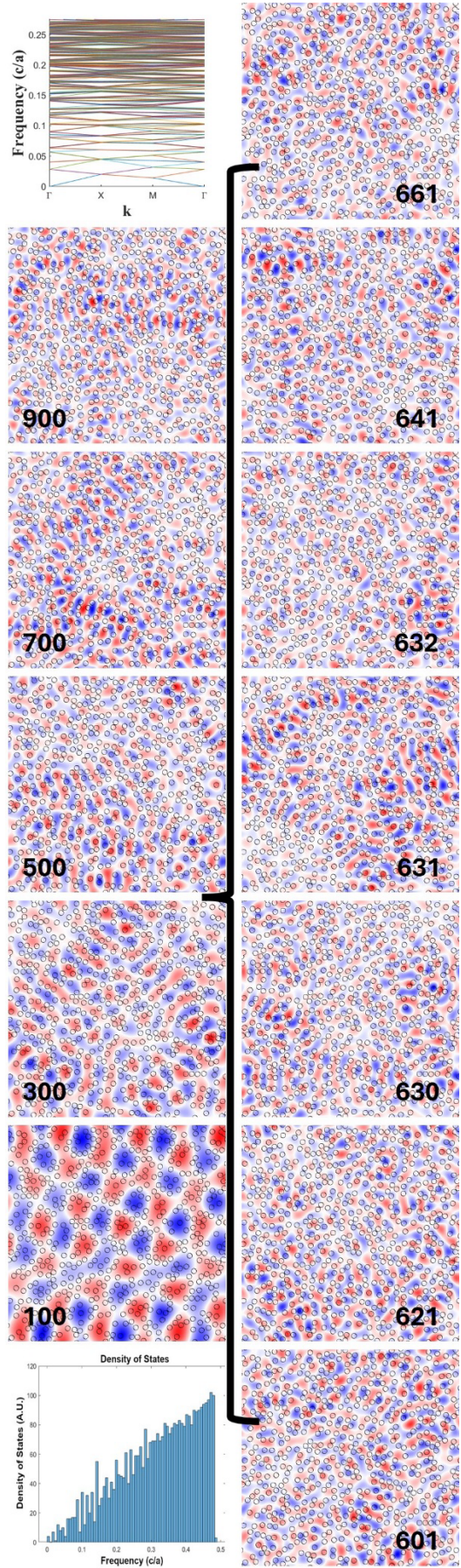

*Figure 2: Simulation results of the band diagram (top left), density of states (bottom left), and electrical field profile of various modes for the extreme case of low dielectric contrast, $\varepsilon_r=2$. The supercell consists of 631 dielectric cylinders. The electrical field profiles of various bands are shown. The positive (negative) values of the electrical field are represented by the red (blue) shade, and the dielectric cylinders are shown in gray circles. The electric field profiles of various bands show similar behavior in this low-dielectric-contrast case, in which a PBG doesn't exist.*

Figure 2 (top left) shows the simulation result of the band diagram using MPB for the case of low dielectric-contrast (2:1). For such a low index-mismatch, we observe that the modes exist for all frequency values under investigation. The histogram of the number of modes as a function of frequency, demonstrating the density of states of the low dielectric-contrast case ($\varepsilon_r=2$), is plotted in Figure 2 (bottom left). When the dielectric contrast is low, the density of states is a continuous function that increases with the frequency and doesn't equal zero at any frequency value. We do not observe any opening of a band gap for which the density of states vanishes.

The E-field profiles of various bands for the low-contrast case are also shown in Figure 2. Red (positive) and blue (negative) shading represent the electrical field strength, while gray circles indicate dielectric cylinders, thus illustrating how the field is distributed relative to the cylinder surfaces and interiors. The corresponding wavelength reduces with the band index as the frequency increases. The E-field profiles for bands 500 to 700 show similar behaviors. From the 601st bands to the 661st bands, we don't observe any differences in E-field profiles. E fields are neither concentrated in the high-index cylinders nor the low-index air regions.

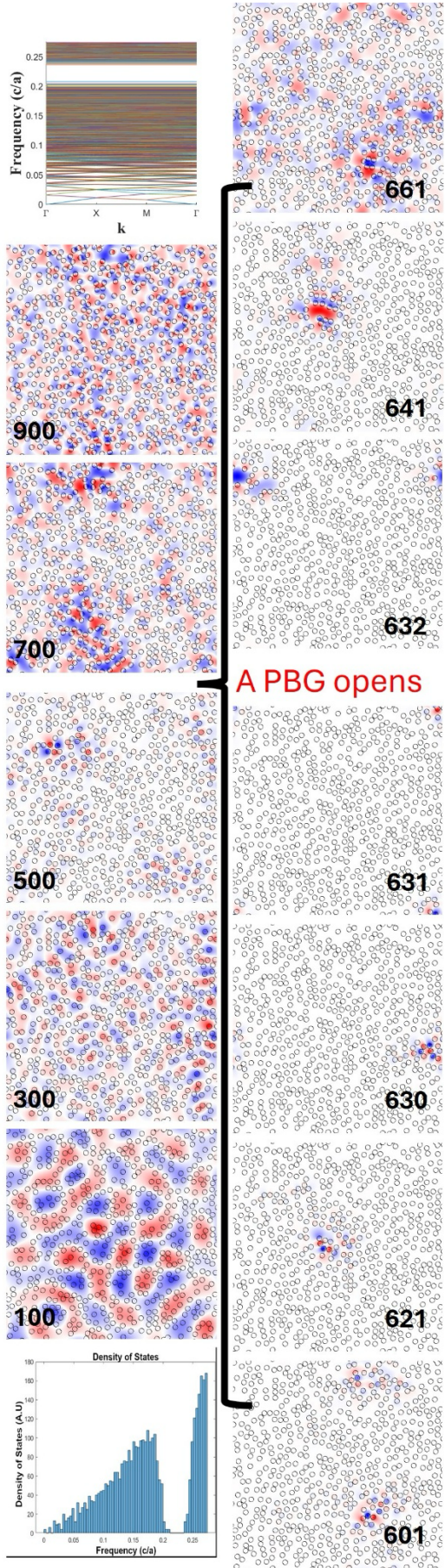

*Figure 3: Simulation results of the band diagram (top left), density of states (bottom left), and electrical field profile of various modes for the extreme case of high dielectric contrasts $\varepsilon_r=12$. The super-cell consists of 631 dielectric cylinders. A wide TM PBG is opened between bands 631 and 632. The positive (negative) values of the field are represented by the red (blue) shade, and the dielectric cylinders are shown in gray circles to demonstrate the relation of the field distribution to the cylinder walls. The electric field profiles of various bands show distinct differences for bands below the gap (concentrated in the dielectric cylinders in Band 631 and lower) or above the bandgap (concentrated in the air in Band 632 and higher) for the high dielectric-contrast case, opening a wide PBG.*

Figure 3 shows the results for the other extreme case of a very high dielectric contrast (12:1). There is a clear bandgap (a forbidden frequency range) shown in the band diagram simulation result (Figure 3 top left). The density of states as a function of frequency (Figure 3 bottom left) shows the existence of a range of frequencies for which the density of states vanishes: light of those frequencies cannot exist in the samples except as evanescent waves, which are associated with imaginary wave vector components [1].

Previous studies [13,15] have shown that, in hyperuniform disordered structures with a high enough dielectric contrast, waves with wavelengths much longer than the average spacing between particles propagate ballistically (i.e., bands 100 and 300 in Figure 3) and waves with much shorter wavelengths propagate diffusively (i. e. bands 700 and 900 in Figure 3) and a photonic bandgap can open in between. Near the edge of the photonic bandgap, the eigenmodes are often localized modes (i.e., Band 500 and bands 601 to 661 in Figure 3).

In the high-dielectric-contrast case, we observe the opening of a wide photonic bandgap for the TM polarization between bands 631 and 632. The histogram of the number of modes as a function of frequency demonstrates the density of states (DoS) (Figure 3 bottom left) for the high contrast case featuring a clear frequency range with zero DoS, the PBG. The E-field profiles for bands between 601 and 661 are shown in Fig. 3. They are localized modes next to the bandgap. Among them, the E-field profiles for bands 630 and lower show localized dielectric bands, whose E-fields are concentrated in the high-index cylinders. Bands 632 to 641 are localized air bands, whose E fields are concentrated in the air regions. Band 631 has the highest eigenvalue of photon energy (frequency) among all 631 dielectric bands.

Note that band 631 (in Figure 3) is a boundary mode introduced by the arbitrary choice of the supercell's boundary in MPB simulations. The periodic boundary condition creates artificially denser or less dense cylinder distributions locally along the boundary of the supercell. These artificial modes may occur inside the actual gap of the original configuration, making the simulated result of the gap appear to be narrower than the true gap of the disordered structure.

More importantly, as shown in Figure 3, in the high-index-mismatch case, for bands 1 to 631, the E-field is concentrated in the high-index cylinders. These are the dielectric bands corresponding to the first band (the dielectric band below the TM PBG) of a regular photonic crystal [8]. Because each supercell contains 631 particles, each band becomes 631 bands for the supercell. On the other hand, for bands 632 and above, the E-field concentrates in the air regions and has nodes in the high-index cylinders. Those are the air bands similar to the air band above the PBG for a photonic crystal. The electric energy concentration factor (CF) is defined by

$$C_F = \frac{\int_{high-index\,region} d^3r \varepsilon(r)|E(r)|^2}{\int_{supercell} d^3r \varepsilon(r)|E(r)|^2} \quad \text{Eq. (1)}$$

and explains the PBG formation in photonic crystals. A significant difference in CFs between the dielectric and air bands of a photonic crystal results in a wider energy gap [8]. For the high-dielectric-contrast case, the CF values for bands 631 and below are much larger than the CF values for bands 632 and above. Hence, given a sufficiently high dielectric contrast, the group of dielectric bands and air bands can be separated completely, similar to how the valence bands and conducting bands are separated for semiconductors. Thus, a bandgap (a range of forbidden frequencies) is opened. The significant difference between the electric field distribution for modes below and above the gap is responsible for the opening of a sizable photonic band gap in hyperuniform disordered structures [15].

On the other hand, for the low-dielectric-contrast case, the CF values for bands below or above band 631 are similar, and their E-field profiles don't show distinct behaviors of dielectric bands or air bands (as shown in Figure 2). Thus, their associated photon energies and eigenvalues are mixed instead of forming two distinct groups to open a bandgap.

3.2 Minimum dielectric contrast for opening the bandgap as a function of the shearing amplitude

Since the existence of a bandgap depends on how high the dielectric contrast $\varepsilon_r$ is, we repeat the above band-diagram simulations using different values of $\varepsilon_r$ to find the minimal dielectric contrast $\varepsilon_{r,min}$, which allows for the opening of a bandgap for each disordered configurations experimentally obtained at different values of the displacement amplitude $\Delta$. The results for $\varepsilon_{r,min}$ as a function of $\Delta/\Delta^*$ are shown in Figure 4. The error bars in Figure 4 are estimated according to the threshold set manually to identify whether a bandgap exists. When the gap is narrow, the artificially introduced modes due to the supercell boundary in the MPB simulations can make the gap less obvious. Typically, when the width of the frequency range with zero density of states is wider than about 2% of the center frequency, we consider it a gap. In that case, the range of zero transmission is also as wide.

Interestingly, we also find two regimes with distinct optical behaviors, corresponding to the two hydrodynamical regimes identified by Weijs *et al.* [17]. A significant difference lies in the value of the minimum dielectric contrast at which the bandgap starts appearing. Near the transition (when $\Delta=0.959\Delta^*$), the opening happens at a dielectric contrast of about 4.0±0.5, whereas above the transition (when $\Delta=1.40\Delta^*$) in the non-uniform regime, the band gap does not open until the dielectric contrast reaches 9.0±0.5 In the regime $\Delta < \Delta^*$, for which the droplets organize in a hyperuniform state, $\varepsilon_{r,min}$ is approximately constant and reaches a plateau value of 4.5±0.5. In the regime $\Delta > \Delta^*$, for which droplets do not organize as uniformly, $\varepsilon_{r,min}$ is also approximately constant at a higher value of 8.2±1.0. The first-order transition, which was already reported and described using the fidelity function [26], translates into a transition in the optical properties with a discontinuity.

A value of dielectric contrast of 4.5 needed for those hyperuniform structures entails a material with a refractive index of about 2.1, which can be reached by replacing the droplets with Cerium Oxide, for example. The dielectric contrast of 8.2 requires a material with an index of refraction of about 2.84, greatly limiting the potential material choices. For example, titanium oxide (index of refraction = 2.87) barely opens a very narrow PBG for those disordered patterns obtained in the regime $\Delta > \Delta^*$, while decorating the disordered patterns obtained with $\Delta$ just below $\Delta^*$ with titanium oxide can provide a much wider gap, whose width is 8.3% of its center frequency.

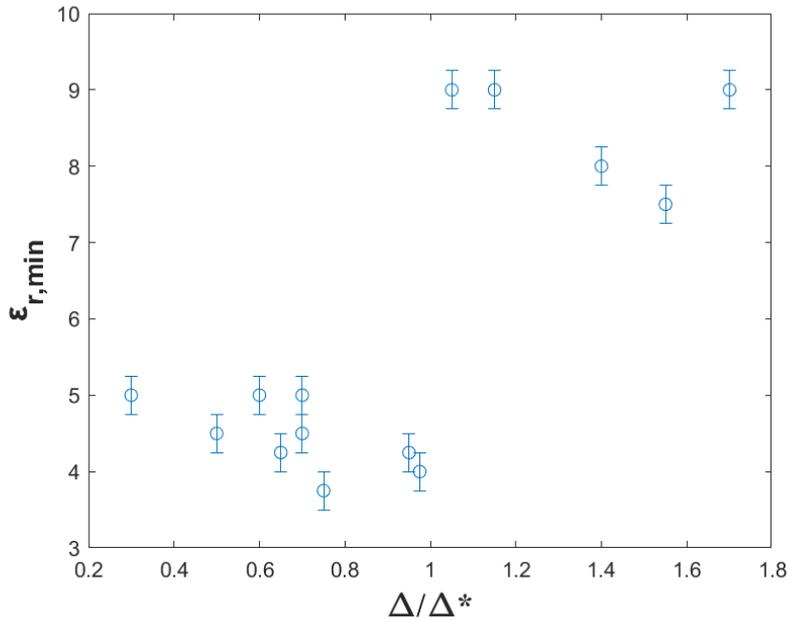

Figure 4: The minimum dielectric contrast to open a PBG for disordered patterns obtained at different displacement amplitudes. $\varepsilon_{r,min}$ is plotted as a function of $\Delta/\Delta^*$. A clear 1$^{st}$ order transition around $\Delta/\Delta^*=1$ is found.

Configurations formed above the transition are much less uniform, with many more large voids and dense cylinder regions than configurations formed below the transition. The associated localized defect modes (i.e., air modes around large voids and dielectric modes at dense cylinder locations) make it significantly harder to open a real PBG with zero density of states. Even when a PBG opens at a high dielectric contrast for those configurations, its width is significantly limited by those localized defect modes near the gap.

In addition, we also use MEEP simulations to calculate the transmission spectrum for the TM polarization through those cylinder-decorated structures by varying $\varepsilon_r$ successively. Figure 5 shows the simulated transmitted intensity as a function of both the frequency and the dielectric contrast for four configurations. Two configurations (Figure 5a and 5b) are obtained below the transition $\Delta^*$, and the other two (Figure 5c and 5d) are obtained far above the transition $\Delta^*$. For all configurations, one can observe the transmission gap, which corresponds to the vanishing transmitted intensity (less than -70dB) in dark blue in Figure 5. Again, we see a significant difference between the cases below and above the transition. For the configurations obtained below the transition (Δ/Δ*=0.775 or 0.959), the transmission gap opens at a much lower minimum dielectric contrast than what is required for configurations formed at Δ>Δ*, For the more uniform structures formed at Δ<Δ* (Figure 5a and 5b), the transmission gap is wide when the dielectric contrast is beyond 5.0.

For the configurations formed above the transition (Δ/Δ*=1.40 or 1.85) (Figure 5c and 5d), the particle distribution is much less uniform and includes many large voids and dense regions; hence, their transmission gaps (-70 dB) are much narrower and open at a much higher dielectric contrast.

The real PBGs with zero density of states found above generally agree with the frequency ranges where the transmission reduces by 70 dB or more in the MEEP simulations. For configurations formed above the transition, the frequency range exhibiting low transmission is broader than the actual PBG, where the density of states is exactly zero. This occurs because, near the edges of the PBG, transmission remains low even at frequencies corresponding to localized defect modes associated with large voids and dense cylinder regions. In those less-uniform configurations formed above the transition, large voids and dense regions are so common that their localized states occupy a noticeable frequency range right above or below the real PBG with zero DoS. For those configurations, the minimum dielectric contrast required for PBG formation (zero DoS) verified in the MPB simulation is even higher than the dielectric contrast at which the transmission first reduces by 70 dB.

This finding clearly demonstrates that while both configurations originate from disordered droplet arrangements, the hyperuniform distribution is more conducive to photonic bandgap formation than the non-hyperuniform one. By controlling the displacement amplitude in the microfluidic channel, large-scale and isotropic hyperuniform patterns can be self-assembled. When the dielectric contrast is large enough to open a PBG in two different structures, increased hyperuniformity leads to a wider photonic bandgap. Hence, we can also modify the width of the gap by controlling the displacement amplitude in the microfluidic channel to create templates with different degrees of hyperuniformity.

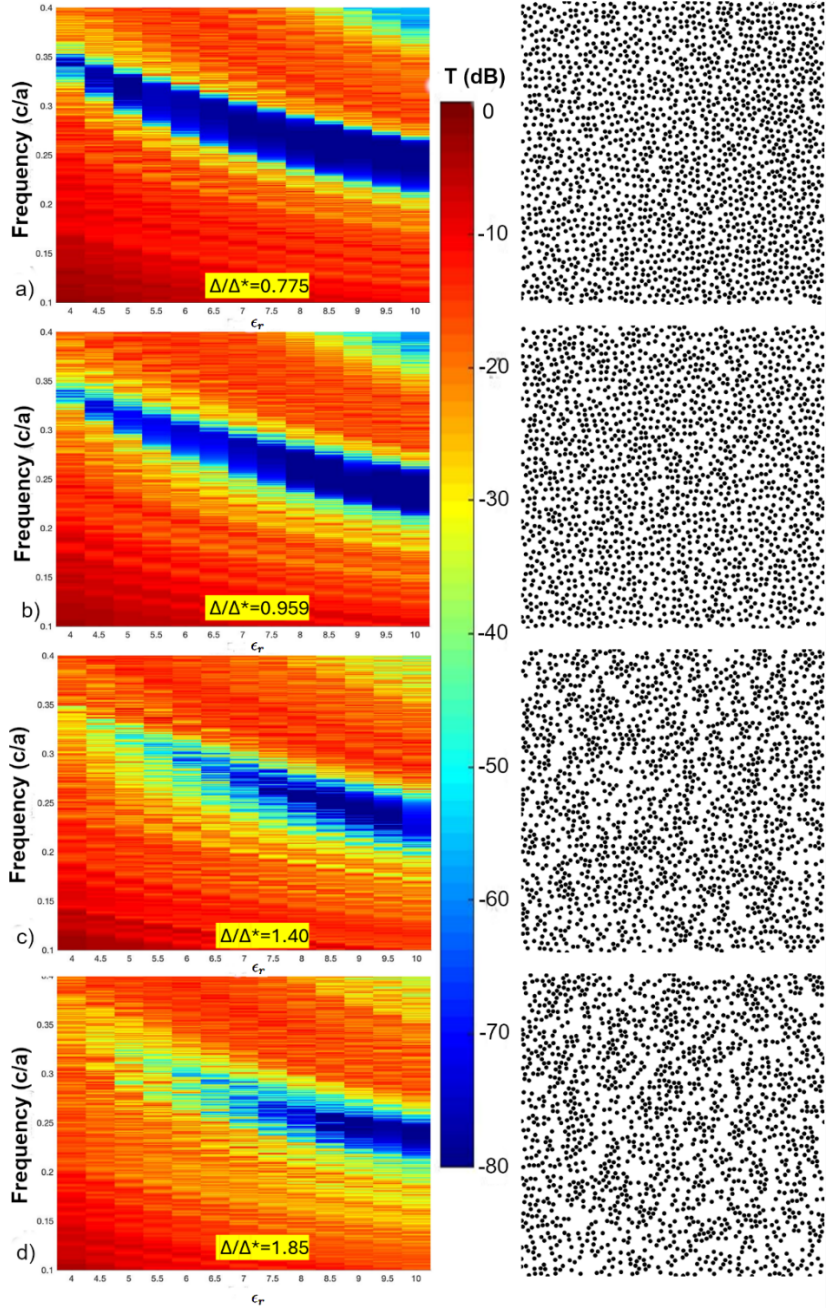

*Figure 5: The MEEP simulation results for transmission (color) through disordered patterns realized with displacement amplitude Δ right below Δ\* (top) and way above Δ\* (bottom) as a function of frequency and the dielectric contrast. When the amplitude Δ is right below the critical amplitude, the resulting pattern is much more uniform and opens a frequency region with -70dB of transmission when the dielectric contrast is 4.0, at which the opening of PBG is verified using the MPB simulation. When the amplitude Δ is above the critical amplitude, the resulting pattern is much less uniform, and a frequency region of -70dB of transmission only opens when the dielectric contrast is beyond 7.0, while MPB simulations suggest the PBG with zero density-of-states only opens at an even higher dielectric contrast (around 8.0).*

4. **Conclusions**

In conclusion, our investigation focused on the optical properties of various disordered configurations derived from hydrodynamics experiments. We started from the patterns formed by an emulsion when subjected to an oscillatory flow. To simulate the optical properties of materials based on these patterns, we assume that the emulsion droplets are further solidified as a template for cylinders with various dielectric constants to grow from. By computing fully vectorial eigenmodes of Maxwell's equations using MPB and conducting FDTD simulations with MEEP, we characterized their photonic bandgaps. Our findings reveal that at the critical amplitude Δ*, the droplets organize into an optimal configuration where the minimum dielectric contrast for opening a photonic bandgap is achieved. This transition occurs right below and above Δ*, suggesting a new method to self-assemble large-scale and isotropic templates for photonic bandgap materials to grow on in a bottom-up approach. Ultimately, this work provides further evidence of how microfluidics can be applied to process novel optical materials [20,32,33].

## Acknowledgments

The authors thank the CRSNG, CNRS, and Solvay for financial support. The authors acknowledge the support of the French National Agency of Research (ANR) for the project REACT through the grant ANR15-PIRE-0001-06. LN2 is an International Research Laboratory (IRL) funded and co-operated by Université de Sherbrooke (UdeS), Centre National de la Recherche Scientifique (CNRS), Ecole Centrale Lyon (ECL), Institut National des Sciences Appliquées de Lyon (INSA Lyon), and Université Grenoble Alpes (UGA). It is also financially supported by the Fond de Recherche du Québec Nature et Technologies FRQNT. We acknowledge the support of the Natural Sciences and Engineering Research Council of Canada (NSERC), *RGPIN-2022-04411, and the support and awards SFSU offers to its PI and students*. We also thank Denis Bartolo and Raphael Jeanneret for sharing their data and for their help in handling it.

## Disclosures

The authors declare no conflicts of interest.

Bibliography

1. J. D. Joannopoulos, S. G. Johnson, J. N. Winn, and R. D. Meade, *Photonics Crystals, Molding the Flow of Light* (2008).
2. N. Muller, J. Haberko, C. Marichy, and F. Scheffold, "Silicon Hyperuniform Disordered Photonic Materials with a Pronounced Gap in the Shortwave Infrared," Adv. Opt. Mater **2**(2), 115–119 (2014).
3. E. Yablonovitch, "Photonic crystals: semiconductors of light," Sci. Am. **285**(6), 46–55 (2001).
4. E. Yablonovitch, T. J. Gmitter, and K. M. Leung, "Photonic band structure: The face-centered-cubic case employing nonspherical atoms.," Phys. Rev. Lett. **67**(17), 2295–2298 (1991).
5. Y. A. Vlasov, X. Z. Bo, J. C. Sturm, and D. J. Norris, "On-chip natural assembly of silicon photonic bandgap crystals.," Nature **414**(6861), 289–293 (2001).
6. A. Imhof and D. J. Pine, "Ordered macroporous materials by emulsion templating," Nature **389**(6654), 948–951 (1997).
7. K. Ishizaki, M. Koumura, K. Suzuki, K. Gondaira, and S. Noda, "Realization of three-dimensional guiding of photons in photonic crystals," Nat. Photonics **7**(2), 133–137 (2013).
8. J. D. Joannopoulos, S. G. Johnson, J. N. Winn, and R. D. Meade, *Photonic Crystals: Molding the Flow of Light* (2008).
9. J. Ricouvier, P. Tabeling, and P. Yazhgur, "Foam as a self-assembling amorphous photonic band gap material.," Proc Natl Acad Sci USA **116**(19), 9202–9207 (2019).
10. M. A. Klatt, P. J. Steinhardt, and S. Torquato, "Phoamtonic designs yield sizeable 3D photonic band gaps.," Proc Natl Acad Sci USA **116**(47), 23480–23486 (2019).
11. S. Torquato and F. H. Stillinger, "Local density fluctuations, hyperuniformity, and order metrics.," Phys. Rev. E Stat. Nonlin. Soft Matter Phys. **68**(4 Pt 1), 041113

(2003).

12. C. E. Zachary and S. Torquato, "Hyperuniformity in point patterns and two-phase random heterogeneous media," J. Stat. Mech. **2009**(12), P12015 (2009).
13. M. Florescu, S. Torquato, and P. J. Steinhardt, "Designer disordered materials with large, complete photonic band gaps.," Proc Natl Acad Sci USA **106**(49), 20658–20663 (2009).
14. W. Man, M. Florescu, E. P. Williamson, Y. He, S. R. Hashemizad, B. Y. C. Leung, D. R. Liner, S. Torquato, P. M. Chaikin, and P. J. Steinhardt, "Isotropic band gaps and freeform waveguides observed in hyperuniform disordered photonic solids.," Proc Natl Acad Sci USA **110**(40), 15886–15891 (2013).
15. W. Man, M. Florescu, K. Matsuyama, P. Yadak, G. Nahal, S. Hashemizad, E. Williamson, P. Steinhardt, S. Torquato, and P. Chaikin, "Photonic band gap in isotropic hyperuniform disordered solids with low dielectric contrast.," Opt. Express **21**(17), 19972–19981 (2013).
16. S. Torquato, G. Zhang, and F. H. Stillinger, "Ensemble theory for stealthy hyperuniform disordered ground states," Phys. Rev. X **5**(2), 021020 (2015).
17. J. H. Weijs, R. Jeanneret, R. Dreyfus, and D. Bartolo, "Emergent hyperuniformity in periodically driven emulsions.," Phys. Rev. Lett. **115**(10), 108301 (2015).
18. R. Dreyfus, Y. Xu, T. Still, L. A. Hough, A. G. Yodh, and S. Torquato, "Diagnosing hyperuniformity in two-dimensional, disordered, jammed packings of soft spheres.," Phys. Rev. E Stat. Nonlin. Soft Matter Phys. **91**(1), 012302 (2015).
19. R. Kurita and E. R. Weeks, "Incompressibility of polydisperse random-close-packed colloidal particles.," Phys. Rev. E Stat. Nonlin. Soft Matter Phys. **84**(3 Pt 1), 030401 (2011).
20. J. Ricouvier, R. Pierrat, R. Carminati, P. Tabeling, and P. Yazhgur, "Optimizing Hyperuniformity in Self-Assembled Bidisperse Emulsions.," Phys. Rev. Lett. **119**(20), 208001 (2017).
21. A. T. Chieco and D. J. Durian, "Quantifying the long-range structure of foams and other cellular patterns with hyperuniformity disorder length spectroscopy.," Phys. Rev. E **103**(6–1), 062609 (2021).
22. A. Chremos and J. F. Douglas, "Particle localization and hyperuniformity of polymer-grafted nanoparticle materials.," Ann. Phys. **529**(5), (2017).
23. S. Wilken, R. E. Guerra, D. J. Pine, and P. M. Chaikin, "Hyperuniform structures formed by shearing colloidal suspensions.," Phys. Rev. Lett. **125**(14), 148001 (2020).
24. D. Hexner, P. M. Chaikin, and D. Levine, "Enhanced hyperuniformity from random reorganization.," Proc Natl Acad Sci USA **114**(17), 4294–4299 (2017).
25. D. Hexner and D. Levine, "Hyperuniformity of critical absorbing states.," Phys. Rev. Lett. **114**(11), 110602 (2015).
26. R. Jeanneret and D. Bartolo, "Geometrically protected reversibility in hydrodynamic Loschmidt-echo experiments.," Nat. Commun. **5**, 3474 (2014).
27. S. Johnson and J. Joannopoulos, "Block-iterative frequency-domain methods for Maxwell's equations in a planewave basis.," Opt. Express **8**(3), 173–190 (2001).
28. A. Taflove and S. C. Hagness, *Computational Electrodynamics The Finite-Difference Time-Domain Method*, 3rd ed. (2005).
29. A. F. Oskooi, D. Roundy, M. Ibanescu, P. Bermel, J. D. Joannopoulos, and S. G. Johnson, "Meep: A flexible free-software package for electromagnetic simulations by the FDTD method," Comput. Phys. Commun. **181**(3), 687–702 (2010).
30. A. T. Chieco, R. Dreyfus, and D. J. Durian, "Characterizing pixel and point patterns with a hyperuniformity disorder length.," Phys. Rev. E **96**(3–1), 032909 (2017).
31. S. R. Sellers, W. Man, S. Sahba, and M. Florescu, "Local self-uniformity in photonic networks.," Nat. Commun. **8**, 14439 (2017).

32. Y. Xu, D. Ge, G. A. Calderon-Ortiz, A. L. Exarhos, C. Bretz, A. Alsayed, D. Kurz, J. M. Kikkawa, R. Dreyfus, S. Yang, and A. G. Yodh, "Highly conductive and transparent coatings from flow-aligned silver nanowires with large electrical and optical anisotropy.," Nanoscale **12**(11), 6438–6448 (2020).
33. F. Bian, L. Sun, L. Cai, Y. Wang, Y. Wang, and Y. Zhao, "Colloidal Crystals from Microfluidics.," Small **16**(9), e1903931 (2020).